\documentclass[journal,10pt]{IEEEtran}
\usepackage{algorithmic}
\usepackage{algorithm}
\usepackage{color}
\usepackage{textcomp}
\usepackage{amsmath,graphicx}
\usepackage{epsfig}
\usepackage{multirow}
\usepackage{csquotes}
\usepackage{array}
\usepackage{makecell}
\usepackage{amssymb}
\usepackage{enumitem}
\usepackage{siunitx}
\DeclareMathOperator*{\argmin}{argmin}

\newcommand*{\argminl}{\argmin\limits}

\usepackage{cite}

\newcommand{\bigO}{\mathcal{O}}

\begin{document}

\title{Estimating Sparsity Level for Enabling Compressive Sensing of Wireless Channels and Spectra in 5G and Beyond}

\author{
Mahmoud~Nazzal,~\IEEEmembership{Member,~IEEE}, Mehmet Ali Ayg\"{u}l,~\IEEEmembership{Student Member,~IEEE,}
and~H\"{u}seyin~Arslan,~\IEEEmembership{Fellow,~IEEE}
\thanks{M.~ Nazzal and M. A.~Ayg\"{u}l are with the Department of Electrical and Electronics Engineering, Istanbul Medipol University, 34810 Istanbul, Turkey (e-mail: mahmoud.nazzal@ieee.org, mehmetali.aygul@ieee.org).
}
\thanks{H.~Arslan is with the Department of Electrical Engineering, University of South Florida, Tampa, FL 33620, USA and also with the Department of Electrical and Electronics Engineering, Istanbul Medipol University, 34810 Istanbul, Turkey (e-mail: arslan@usf.edu).}
\thanks{This work has been submitted to the IEEE for possible publication. Copyright may be transferred without notice, after which this version may no longer be accessible.}}
\maketitle

\begin{abstract}
Applying compressive sensing (CS) allows for sub-Nyquist sampling in several application areas in 5G and beyond. This reduces the associated training, feedback, and computation overheads in many applications. However, the applicability of CS relies on the validity of a signal sparsity assumption and knowing the exact sparsity level. It is customary to assume a foreknown sparsity level. Still, this assumption is not valid in practice, especially when applying learned dictionaries as sparsifying transforms. The problem is more strongly pronounced with multidimensional sparsity. In this paper, we propose an algorithm for estimating the composite sparsity lying in multiple domains defined by learned dictionaries. The proposed algorithm estimates the sparsity level over a dictionary by inferring it from its counterpart with respect to a compact discrete Fourier basis. This inference is achieved by a machine learning prediction. This setting learns the intrinsic relationship between the columns of a dictionary and those of such a fixed basis. The proposed algorithm is applied to estimating sparsity levels in wireless channels, and in cognitive radio spectra. Extensive simulations validate a high quality of sparsity estimation leading to performances very close to the impractical case of assuming known sparsity. 
\end{abstract}


\IEEEpeerreviewmaketitle

\section{Introduction}

\par In its primitive definition, the sparsity of a signal (or a system) is its admittance to concentrating most of its energy in a few coefficients \cite{pastor2015mathematics}. Said conversely, a sparse signal has a majority of negligibly small coefficients. This intrinsic property naturally appears in many application areas. Similar to its appeal in these areas, sparsity has been receiving attention for applications in communications of the fifth-generation (5G) and beyond. An immediate advantage of sparsity is the applicability of the recently brought concept of compressive sensing (CS) \cite{Compressive_sesing}. CS allows for sub-Nyquist sampling, i.e., sampling and reconstructing a few influential coefficients, rather than all the coefficients of a signal. This promises to significantly reduce the underlying sampling, training, feedback, and computational costs in many applications in 5G and beyond \cite{choi2017compressed}.

\par Two major aspects of sparsity in 5G and beyond are those of the channel and spectrum \cite{choi2017compressed,qin2018sparse}. From a channel estimation perspective, pilot training and feedback have remained a bottleneck against efficient channel estimation. Fortunately, channel sparsity enables sub-Nyquist channel estimation. This allows for reduced pilot usage and reformulates channel estimation as a sparse recovery operation. On the other hand, a major challenge in spectrum sensing (SS) for cognitive radio (CR) is the need for ultra-fast analog-to-digital converters (ADCs) to sample the (wideband) spectrum and estimate its occupancies. In this regard, exploiting spectrum sparsity enables sub-Nyquist SS, thereby reducing the hard requirements on ADCs. Therefore, this paper highlights sparsity level estimation in channel and spectrum estimation as two case studies.

\par Recently, there has been an increasing belief in the existence of channel sparsity in 5G and beyond. Furthermore, it is believed to exist in multiple domains (dimensions), such as the angle, delay, and Doppler frequency domains \cite{nazzal2020channel}. A well-known example of a sparse channel is in the massive multiple input multiple output (mMIMO) case. A typical sparsity aspect is an angle-domain sparsity motivated by the limited scattering and narrow angular spread in practical mMIMO systems \cite{choi2017compressed}. Besides, employing many antennas the base station (BS) improves spatial resolvability thereby promoting sparsity. Furthermore, carrier frequencies in 5G and beyond are expected to go higher. This limits the number of effective propagation paths by increasing wave scattering and absorption, as in the case of millimeter-wave (mmWave) communications \cite{sayeed2002deconstructing}. Sparsity domains include also the delay and Doppler-domains where sparsity arises when the domain spread is large compared to the number of dominant propagation paths.

\par Wireless channels in other beyond 5G technologies are also reported to exhibit sparsity. Examples include mmWave and terahertz (THz) communications. Channels in the hybrid technology of mmWave mMIMO are known to be sparse in the beamspace \cite{brady2013beamspace}. Furthermore, the anticipated THz technology for beyond 5G is known to promote channel sparsity \cite{zhang2020channel}. Similar to the mmWave case, THz waves have strong molecular absorption and reflection loss enhancing channel sparsity.

\par The underutilization of the spectrum renders its frequency domain sparsity \cite{sun2013wideband}. Moreover, spectrum sparsity can be exhibited in other domains. The sub-Nyquist SS paradigm is based on assuming spectrum sparsity in a certain domain \cite{davenport2010signal}. There has been a rich literature on the exploitation of CS as a framework for SS. The main outcome is reducing the number of sampling points in sensing the spectrum.

\par In the CS paradigm, knowing the exact sparsity level of the signal is of great importance. This quantity dictates the optimal number of compressive measurements in the sensing stage and controls the quality of the sparse estimation in the reconstruction stage. The de facto in the CS literature is to assume a known sparsity level. This assumption has been widely believed to form a significant gap between the CS theory and its applicability \cite{lopes2016unknown}. Improperly assumed sparsity level results in either early or late termination of sparse recovery algorithms. This can lead to insufficient signal sampling and noise domination, respectively. Therefore, the CS literature, particularly for channel \cite{qin2018sparse} and spectrum \cite{sharma2015cognitive} estimation, necessities the need for efficient, accurate, and timely identification of the sparsity level of those quantities. In a problem such as channel estimation, addressing multidimensional sparsity, this need is more strongly evident \cite{choi2017compressed}. As an alternative to assuming known sparsity, some approaches assume knowing the received signal noise floor. From a practical perspective, this information is neither available in practice.

\subsection{Motivation and Related Works}

\par There are two broad approaches to dealing with unknown sparsity levels. First, is the sequential sampling approach \cite{malioutov2010sequential, malioutov2008compressed,van2018joint,sun2012adaptive}, where one samples the signal with increasing sparsity until a certain stopping criterion is met. These methods require solving a minimization problem at each newly acquired measurement and tend to be too complex. Besides, they lack robustness against noise. The second broad approach is the estimation approach \cite{wang2012sparsity, lopes2016unknown}, where the sparsity level is estimated before applying CS. Along this line, cross-validation (CV) has been employed \cite{ward2009compressed,malioutov2008compressed}. CV is a model-order-estimation technique. In this regard, CV is based on choosing an optimal sparsity level from a set of possible values, where each value corresponds to a certain empirical error. However, the computational cost of this method remains prohibitively high.

\par Within the estimation approach, a research trend considers inferring sparsity level taking advantage of specially designed sensing matrices tailored for this purpose. However, this puts certain limitations on the design of these matrices which may contradict their function of optimizing the sensing. Along this line, Cauchy and Gaussian distributed measurement matrices give rise to a numerical measure of sparsity in \cite{lopes2013estimating}. This measure is quantified by the ratio of squared $\ell_{1}$ and $\ell_{2}$ norms of the signal. However, this relies on knowing the noise variance to adjust one of the parameters in the sensing matrix distribution. This variance is not known in practice. Therefore, as a subsequent improvement, \cite{lopes2016unknown} considers using a family of entropy-based sparsity measures based on limiting distributions. Still, this approach shares the same limitations as the one in \cite{lopes2013estimating}. Another work infers signal sparsity from the sparsity of the sensing process itself \cite{ravazzi2016signal}. Despise its computational efficiency, this approach performs well only in highly-dimensional settings.

\par Another source of inferring the sparsity level is linking it to signal rank estimation. Examples include the multiple measurement vector problem \cite{sharma2014compressive}, in stationary environments \cite{lavrenko2015sparsity} and block-stationary signals \cite{lavrenko2015detection}. However, these approaches assume stationarity in the support and temporal variations of the sparse representation coefficients which are not necessarily valid. Another work, in the SS context, considers roughly inferring spectrum usage sparsity level from geolocation data provided \cite{qin2015data}. Despite its advantages, this method relies on the availability of timely and correct geolocation data and performs poorly if such data is incorrect or unavailable. In summary, the existing methods of sparsity level estimation suffer from experimental setup dependency \cite{wang2012sparsity}, affecting the sensing quality, and high computational and time costs \cite{van2018joint}.

\subsection{Contributions, Notation, and Organization}

\par In view of the above discussion, we propose an algorithm for timely and efficient multidimensional sparsity level estimation and exploitation. Here is a brief account of the contributions in this paper.
\begin{itemize}[leftmargin=*]
\item \textit{Sparsity level estimation}: proposing an algorithm for estimating the sparsity level of the channel/spectrum with respect to a learned dictionary using machine learning (ML). This algorithm is based on performing an easy calculation of the sparsity level over a compact fixed basis function, and then using machine learning to infer the sparsity level over the dictionary. 
\item \textit{Dictionary learning addressing multidimensional sparsity}: proposing dictionary learning as a means of obtaining the sparsifying transform that leads to harvesting multiple sparsity aspects at the same time. 
\end{itemize}

\par Extensive experiments are presented to validate the above propositions. The proposed sparsity estimation algorithm is shown to reliably and efficiently predict the underlying sparsity level in both channel and spectrum estimation settings. In particular, experiments conducted in the channel estimation scenario show a considerable improvement due to the exploitation of multidimensional sparsity.

\par \textit{Notation:} Lower-case plain-faced, lower-case bold-faced and upper-case bold-faced letters represent scalars, vectors, and matrices, respectively. In a matrix $\boldsymbol{X}$, the symbol $\boldsymbol{X}_i$ denotes its $i\mbox{-}$th column. $\mathbb{C}$ denotes the complex number field. The Moore-Penrose pseudoinverse operator is denoted by $\dagger$. The symbols ${\|.\|}_2$ and $\|.\|_0$ represent the $\ell_{2}$ norm and the number of nonzero elements in a vector, respectively.

\par \textit{Organization:} This paper is organized as follows. Section \ref{Section2} revises the preliminaries. The proposed contributions are detailed in Section \ref{Section3}. Section \ref{Secion4} presents simulations and experiments, with the conclusions made in Section \ref{Section5}.

\section{Preliminaries}
\label{Section2}

\subsection{Compressive Sensing}

\par Let $\boldsymbol{y} \in \mathbb{C}^n$ denote a vector signal. The notion of CS considers obtaining a compressed measurement $\boldsymbol{y}_c=\boldsymbol{\Phi}\boldsymbol{y}$ where $\boldsymbol{\Phi} \in \mathbb{C}^{m\times n} $ is a measurement/sensing matrix, with $m< n$, rather than measuring every element in $\boldsymbol{y}$. Clearly, an $n$-to-$m$ dimensionality reduction is made possible by this undersampling operation. It is noted that CS is only applicable to compressible signals; those being sparse explicitly, or have a sparse representation in a certain domain \cite{davenport2010signal}. Since $\boldsymbol{y}$ is not necessarily sparse in its own shape, its sparse representation is typically obtained using a sparsifying transform/basis ($\boldsymbol{\Psi}$); either a fixed basis or a redundant (overcomplete) learned dictionary. In this context, the signal can be approximated as $\boldsymbol{y}=\boldsymbol{\Psi}\boldsymbol{w}$, where $\boldsymbol{w}$ is a sparse coding coefficient vector having only $s \ll n$ nonzero elements. Obtaining $\boldsymbol{w}$ from $\boldsymbol{y}_c$ can be formulated as follows 
\begin{equation}
\argminl_{\boldsymbol{w}} {\|\boldsymbol{w}\|}_0 ~ s.t. ~ \boldsymbol{y}_c=\boldsymbol{\Phi}\boldsymbol{y}= \boldsymbol{\Phi}\boldsymbol{\Psi}\boldsymbol{w}.\label{eq1}
\end{equation}

\par The inverse problem in (\ref{eq1}) is inherently ill-posed. Still, the sparsity of the solution lends itself as an efficient regularizer to this problem under mild conditions. In this regard, the restricted isometry property (RIP) \cite{Compressive_sesing} of $\boldsymbol{\Phi}$ assures a unique solution with high probability. Besides, a number of compressed measurements $m$ being at least equal to $(cs \log n/m)$ for some small constant $c > 0$ assures exact recovery according to the robust uncertainty principle \cite{Compressive_sesing}. Technically, a variety of sparse recovery techniques can be applied to obtain $\boldsymbol{w}$ given $\boldsymbol{y}_c$, $\boldsymbol{\Phi}$ and $\boldsymbol{\Psi}$. To this end, the fundamental intuition behind CS is measuring only the nonzero elements in $\boldsymbol{w}$. Hence, it resembles a compressed measurement of the original signal. Finally, the original signal can be reconstructed as $\hat{\boldsymbol{y}}=\boldsymbol{\Psi}\boldsymbol{w}$.

\subsection{Dictionary Learning for Sparse Coding}

\par A natural signal $\boldsymbol{y} \in \mathbb{C}^n$ is typically compressible, i.e., has a sparse representation $\boldsymbol{w} \in \mathbb{C}^k$ in a specific domain reached by a sparsifying transform $ \boldsymbol{D} \in \mathbb{C}^{n \times k}$. Thus, $\boldsymbol{y}$ can be approximated as $\boldsymbol{y}\approx\boldsymbol{Dw}$. For a given $\boldsymbol{y}$ and $\boldsymbol{D}$, the problem of finding $\boldsymbol{w}$ is referred to as \textit{sparse representation}, formally expressed as
\begin{equation}
\operatorname*{arg\,min}_{\boldsymbol{w}} {\|\boldsymbol{w}\|}_0 ~ s.t. ~ {\|\boldsymbol{y}-\boldsymbol{Dw}\|}_2^2 <\epsilon,\label{eq111}
\end{equation}
\noindent where $\epsilon$ is an error tolerance. Despite its computationally demanding nature, an algorithm such as the orthogonal matching pursuit (OMP) \cite{pati1993orthogonal} gives efficient approximate solutions.

\par One may train for a sparsifying dictionary\footnote{A learned dictionary is usually redundant, i.e., has more columns than rows. The column dimension, i.e., the number of rows, is equal to the signal space dimension.} over a set of training data signals $\boldsymbol{Y} \in \mathbb{C}^{n\times l}$. This is referred to as the \textit{dictionary learning} process \cite{starck2015sparse}, which can be mathematically expressed as 
\begin{equation}
\argminl_{\boldsymbol{W, D}} {\|\boldsymbol{W}_i\|}_0 ~ s.t. ~ {\|\boldsymbol{Y}_i- \boldsymbol{DW}_i\|}_2^2 < \epsilon, ~\forall~ ~i=1, \ldots l.\label{eq3}
\end{equation}

\subsection{Machine learning}

\par The successful application of ML models in diverse application areas ranging from pattern recognition to image and speech processing motivated their applicability to the area of wireless communication \cite{jiang2016machine}. In general, ML approaches can be split into three main categories based on their learning methodology as supervised, unsupervised, and reinforcement learning. Supervised algorithms are widely used for classification and regression problems when a labeled dataset is available. Although the purpose of this work is to obtain an integer value (limited class) we consider this problem as a regression problem. This is because the sparsity level may have many values especially in multidimensions and thus it may require many classes, and the number of classes is not readily available in practice. On the other hand, the regression problem estimates a numeric value and it is not necessarily an integer. Since the sparsity level is an integer, we round the obtained numerical value to the closest integer.

\par Amongst supervised learning models, the feed-forward model has received growing attention for regression problems since it can find numerical values accurately and quickly \cite{kwok1995constructive}. This model can be used with a single-hidden layer for simple problems. On the other hand, the number of neurons can be selected considering the generalizability of the model. Generally speaking, employing more neurons can improve the learning of the dataset. However, this comes at the cost of higher computation and the risk of potentially overfitting the model. In this context, model overfitting refers to the case where the model performs perfectly with the data over which it is trained, but poorly on new (unforeseen) data samples during its usage. Also, other hyper-parameters such as the learning rate can be adjusted to design optimum models.

\section{Estimating the Unknown Multidimensional Sparsity in 5G and Beyond}
\label{Section3}

\subsection{System Model}
\par
\begin{figure}[!t]
\centering
\resizebox{0.9\columnwidth}{!}{
\includegraphics{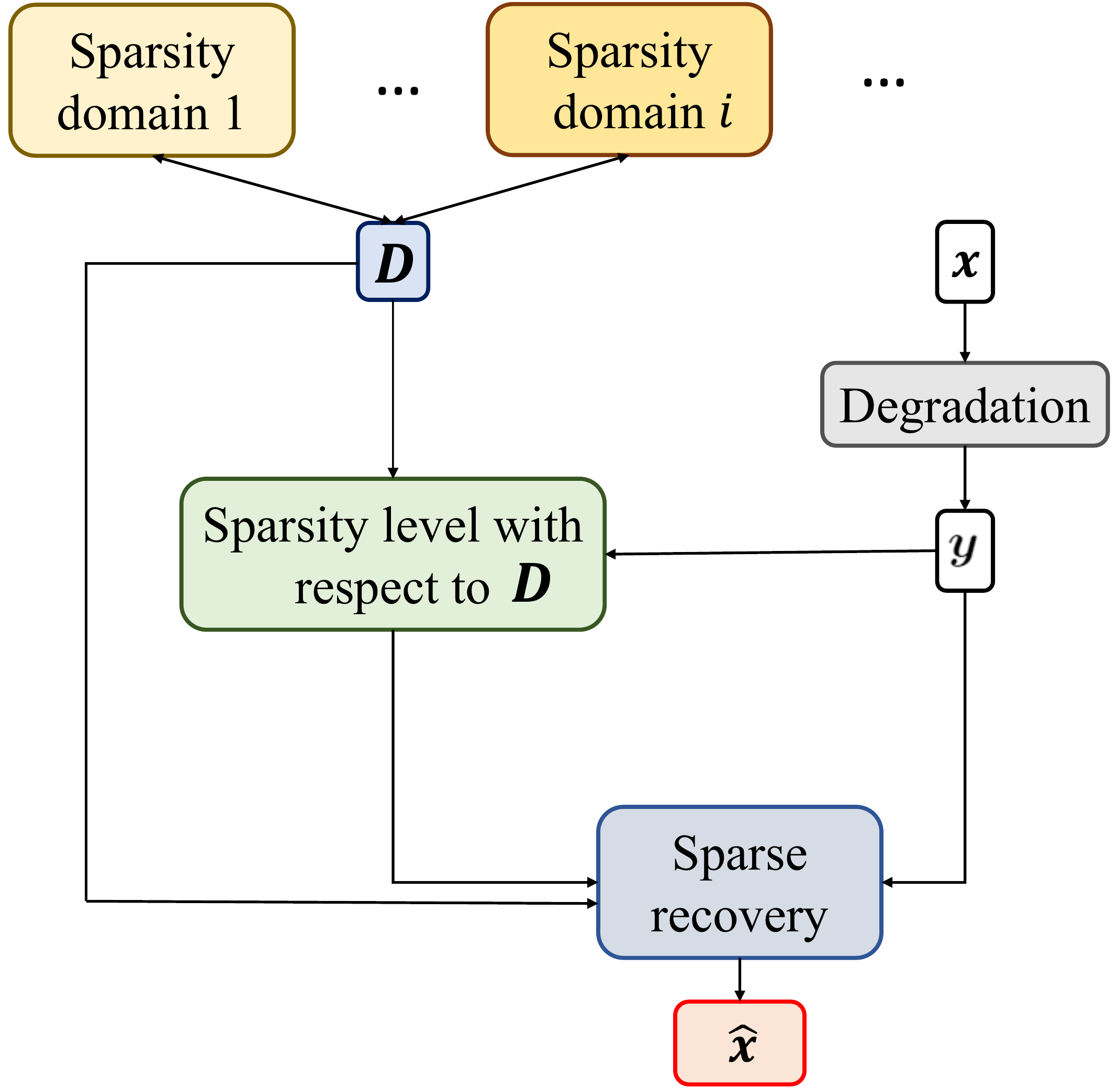}}
\linespread{1}
\caption{The system model for sparsity level estimation.}
\label{system_model}
\end{figure}

 This work considers estimating and exploiting the sparsity of a signal or a system $\boldsymbol{x}$, depending on a distorted measurement $\boldsymbol{y}$, commonly observed at a receiving end, as
\begin{equation}
 \boldsymbol{y}=\boldsymbol{Ax}+\boldsymbol{n}
\end{equation}
\noindent where $\boldsymbol{A}$ denotes a sampling operator and $\boldsymbol{n}$ is additive-white Gaussian noise. It is assumed that $\boldsymbol{x}$ is sparse in a domain defined by a dictionary $\boldsymbol{D}$. So, the aim is to first estimate the sparsity level of $\boldsymbol{x}$ over $\boldsymbol{D}$, based only on $\boldsymbol{y}$ since it is the available observation of the unknown $\boldsymbol{x}$. The eventual aim is to obtain an estimate $\boldsymbol{\hat{x}}=\boldsymbol{D}\boldsymbol{w}$, where $\boldsymbol{w}$ is the sparse coding of $\boldsymbol{y}$ over $\boldsymbol{D}$ according to the estimated sparsity level. This estimation allows for CS and its corresponding sparse recovery. Fig. \ref{system_model} illustrates this system model. This system model applies to both channel and spectrum estimation settings.

\subsubsection{Channel Estimation}

\par As a case study, we consider a MIMO-OFDM system with $K\mbox{-}$subcarriers where the BS has $N$ antennas, while each user has $n$ antennas. A uniform linear array (ULA) is considered at the BS. Message data is mixed with known training pilots at prescribed locations at the transmitter end. After that, data and pilot symbols are subject to an inverse fast Fourier transform stage to yield the corresponding time-domain signal, before adding a cyclic prefix (CP) to it and transmitting it over the channel. At the receiver end, the CP is removed and a Fourier transform operation is carried out to obtain the frequency-domain signal. Next, the channel is estimated according to the known pilots. Eventually, the channel at non-pilot locations is obtained by sparse recovery mimicking an interpolation operation. As a channel model, we use two different models; the geometry-based stochastic channel model (GSCM) and virtual channel model (VCM). The GSCM model is widely used to express MIMO channel modeling as a benchmark while VCM can incorporate channel sparsity in the multiple domains.

\par The channel, $\boldsymbol{h}$, between the BS and any node with a single antenna can be expressed with GSCM \cite{GSCM} as follows
\begin{equation}
\boldsymbol{h}=\sum_{i=1}^{N_c} \sum_{j=1}^{N_s} \boldsymbol{\alpha}_{ij} \boldsymbol{\beta}(\theta_{ij}),\label{equa69}
\end{equation}
\noindent where $N_c$ and $N_s$ represent the number of scattering clusters and the number of sub-paths in each cluster, respectively, and $\boldsymbol{\alpha}_{ij}$ denotes the complex gain of the $j\mbox{-}$th propagation subpath in the $i\mbox{-}$th scattering cluster. The normalized array response at the UE can be represented by $\boldsymbol{\beta}(\theta_{ij})$, where the angle of arrival/departure of the $j\mbox{-}$th subpath in the $i\mbox{-}$th scattering cluster is denoted by $\theta_{ij}$.

\par To incorporate sparsity in the angle, delay, and Doppler spread domains, we use the well-known virtual VCM \cite{bajwa2010compressed}. According to the VCM, the channel frequency response (CFR) at time $t$ is given by
\begin{equation} \label{vcm}
\begin{split}
{\boldsymbol H}(t,f)=&\, \sum_{a=1}^{N_{R}}\sum_{b=1}^{N_{T}}\sum_{c=0}^{L-1}\sum_{d=-M}^{M}\!\!H_{v}(a,b,c,d){\bf a}_{R}\left({\frac{a}{N_{R}}}\right)\cr \\
&\times{\bf a}_{T}^{\rm H}\left(\frac{b}{N_{T}}\right)e^{-j2\pi{\frac{c}{W}}f}e^{j2\pi{\frac{d}{T}}t},
\end{split}
\end{equation}
\begin{equation} 
\begin{split}
H_v(a,b,c,d)\approx&\,\sum_{n\in S_{R,a}\cap S_{T,b}\cap S_{\tau,c}\cap S_{\nu,d}}\!\!\beta_{n}f_{N_{R}}(a/N_{R}-\theta_{R,n})\cr&\times f_{N_{T}}^{\ast}(b/N_{T}-\theta_{T,n})\cr& \times{\rm sinc}(d-T\nu_{n},c-W\tau_{n}),
\end{split} \label{eq77}
\end{equation}
\noindent where $N_R$, $N_T$, $L$, $M$, and $\beta_n$ denote the maximum numbers of resolvable angles of arrival (AoA)s, angles of departure (AoD)s, delays and Doppler shifts within the channel spreads, and the is the complex path gain, respectively. We denote by $T$ the symbol duration and by $W$ the (two-sided) bandwidth. The symbols ${\bf a}_{T}(\frac{b}{N_{T}})$ and ${\bf a}_{R}(\frac{a}{N_{R}})$ represent the array steering and response vectors, respectively.

\par In (\ref{eq77}), the virtual channel coefficients ${H_v(a,b,c,d)}$ are presented. A virtual channel coefficient is approximately equal to the sum of the complex gains of all physical paths whose angles, delays and Doppler shifts lie within an angle-delay-Doppler resolution bin of size $\bigtriangleup{\theta_R}\times\bigtriangleup{\theta_T}\times\bigtriangleup\tau\times\bigtriangleup v$ centered around the virtual sample point $(\theta_R,\theta_T,\tau,\nu)=(a/N_R,b/N_T,c/W,d/T)$ in the angle-delay-Doppler domain. Here, $\nu$ and $\tau$ represent the Doppler spread and delay spreads, respectively, and $N$ denotes the number of propagation subpaths in each resolution bin. The smoothing kernels $f_{N_{R}}(\theta_T)$ and $f_{N_{T}}(\theta_T)$ are Dirichlet kernels, $f_N(\theta)=(1/N)\sum_{a=0}^{N-1}e^{-j2\pi{i}\theta}$, while the two-dimensional sinc kernel identified by $sinc(x,y)=e^{-j\pi{x}}sin(\pi x)sin(\pi y)/({\pi}^2xy)$.

\par Sampling (\ref{vcm}) at $t = iT$ and $f = j\Delta f$ where $\Delta f = 1/T$ is the subcarrier spacing, and with proper CP and timing, the CFR can be written as 
\begin{equation} \label{eq7}
\begin{split}
\boldsymbol{h}[i,j]=&\,\sum_{a=1}^{N_{R}}\sum_{b=1}^{N_{T}}\sum_{c=0}^{L-1}\sum_{d=-M}^{M}\!\!H_{v}(a,b,c,d){\bf a}_{R}\left({\frac{a}{N_{R}}}\right)\cr \\
&\times{\bf a}_{T}^{\rm H}\left(\frac{b}{N_{T}}\right)e^{-j2\pi{\frac{c}{W}}j\Delta f}e^{j2\pi{\frac{d}{T}}iT.} 
\end{split}
\end{equation}
\par Next, a Fourier transformation of the CIR appearing in (\ref{equa69}) or (\ref{eq7}) is given as 
\begin{equation}
\boldsymbol{H}= \boldsymbol{Fh}, \label{eq8}
\end{equation}
\noindent where $\boldsymbol{H}$ has as its columns the channel at each subcarrier, $\boldsymbol{F}$ is the unitary fast Fourier transform (FFT) matrix, and $\boldsymbol{h}$ has as its columns the channel impulse response (CIR) vectors.

\par The received signal at the $i\mbox{-}$th subcarrier of the $j\mbox{-}$th OFDM block after CP removal and FFT transform can be expressed as \cite{jeon2000efficient}
\begin{equation}
\boldsymbol{Y}[i, j]=\boldsymbol{X}[i, j]\boldsymbol{H}[i, j]+\boldsymbol{N}[i, j], \label{eq9}
\end{equation}
\noindent where $\boldsymbol{N}[i,j]$ denotes zero-mean complex additive white Gaussian noise (AWGN) at the $i\mbox{-}$th subcarrier of the $j\mbox{-}$th block. Now, the purpose of channel estimation is to obtain an estimate $\boldsymbol{\hat{h}}$ with the knowledge of the pilot symbols and their respective locations. Next, a channel estimate $\boldsymbol{\hat{h}}$ is obtained as the inverse FFT of $\boldsymbol{\hat{h}}$.

\subsubsection{Spectrum Sensing}

\par In a CR setting, the spectrum is known to be sparse in the frequency domain. It is of interest to estimate the sparsity level of the spectrum. In this regard, we assume dictionary-defined sparsity which may be in multiple domains defined by the dictionary. So, the objective is to estimate the sparsity of the spectrum. Spectrum occupancy means having a primary user (PU) signal, whereas its absence characterizes a spectrum hole. These possibilities correspond to the following hypotheses ($\mathcal{H}_0$) and ($\mathcal{H}_1$), as follows
\begin{equation}
\boldsymbol{y}=
\left\{\begin{array}{ll}
\boldsymbol{n}, & \mathcal{H}_0: \text{there is no PU} \\
\boldsymbol{Hx}+\boldsymbol{n}, & \mathcal{H}_1 : \text{a PU is present},\label{eq21}
\end{array} \right.
\end{equation}
\noindent where $\boldsymbol{H}$, $\boldsymbol{x}$, and $\boldsymbol{n}$ represent the channel matrix, the PU signal, and noise, respectively. In this context, the sparsity level of the spectrum is the number of occupied bands.

\subsection{Learned Dictionaries for Exposing Multidimensional Sparsity}

\par
\begin{figure}[!b]
\centering
\begin{tabular}{@{}c@{}}
\includegraphics[width=.47\linewidth]{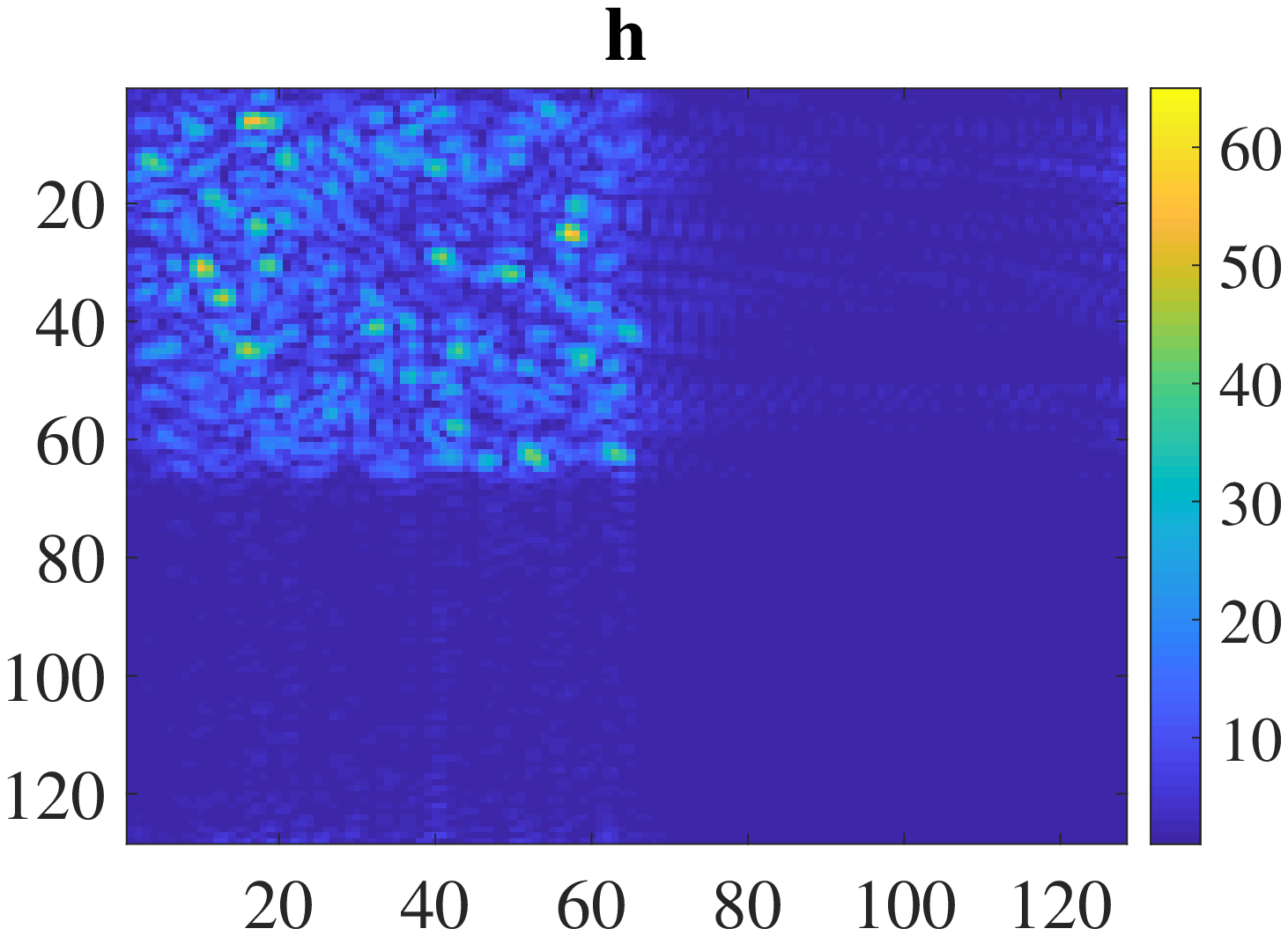} \\
\small (a) 
\end{tabular}
\begin{tabular}{c}
\includegraphics[width=.47\linewidth]{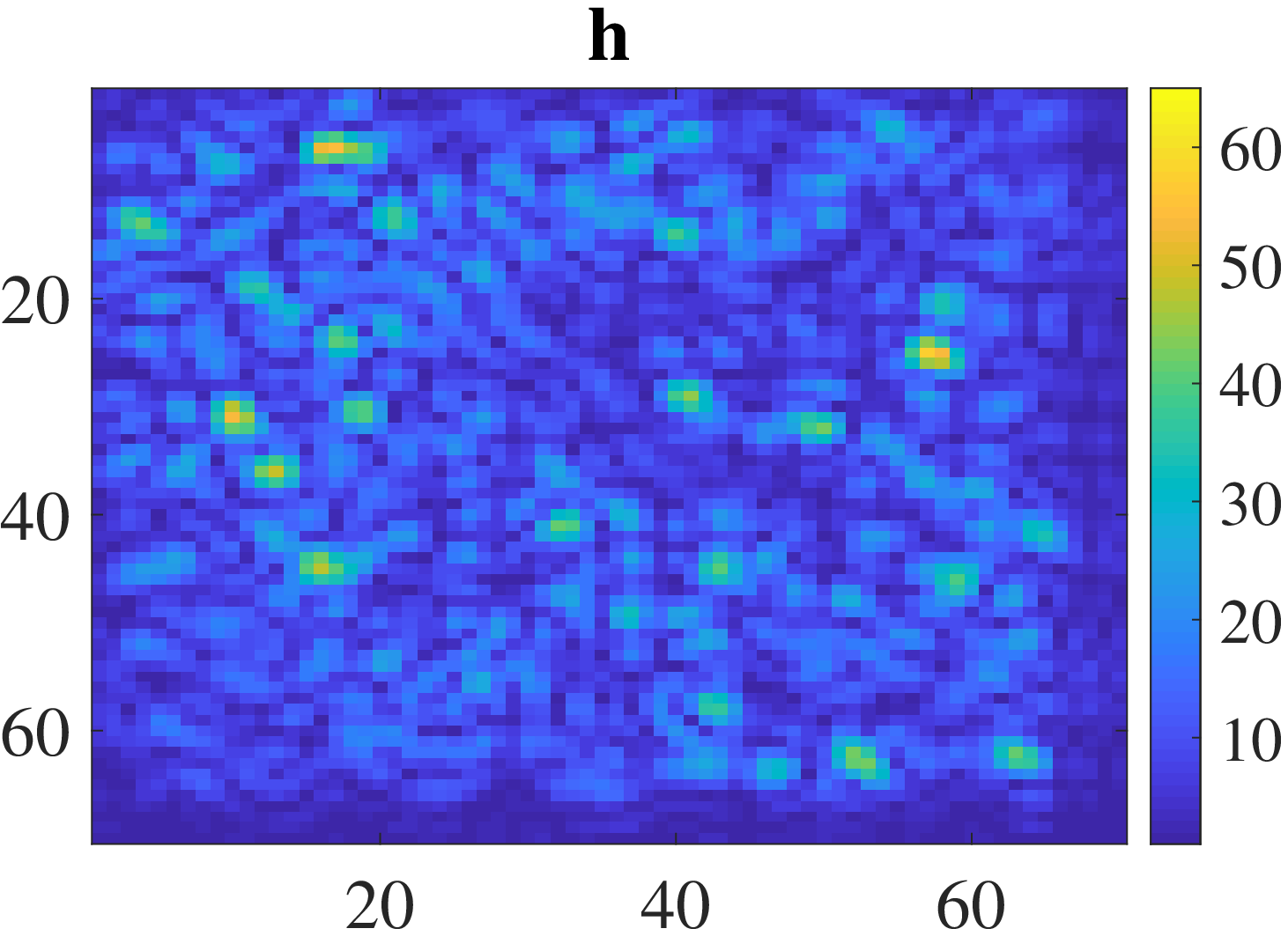} \\
\small (d) 
\end{tabular}
\begin{tabular}{@{}c@{}}
\includegraphics[width=.47\linewidth]{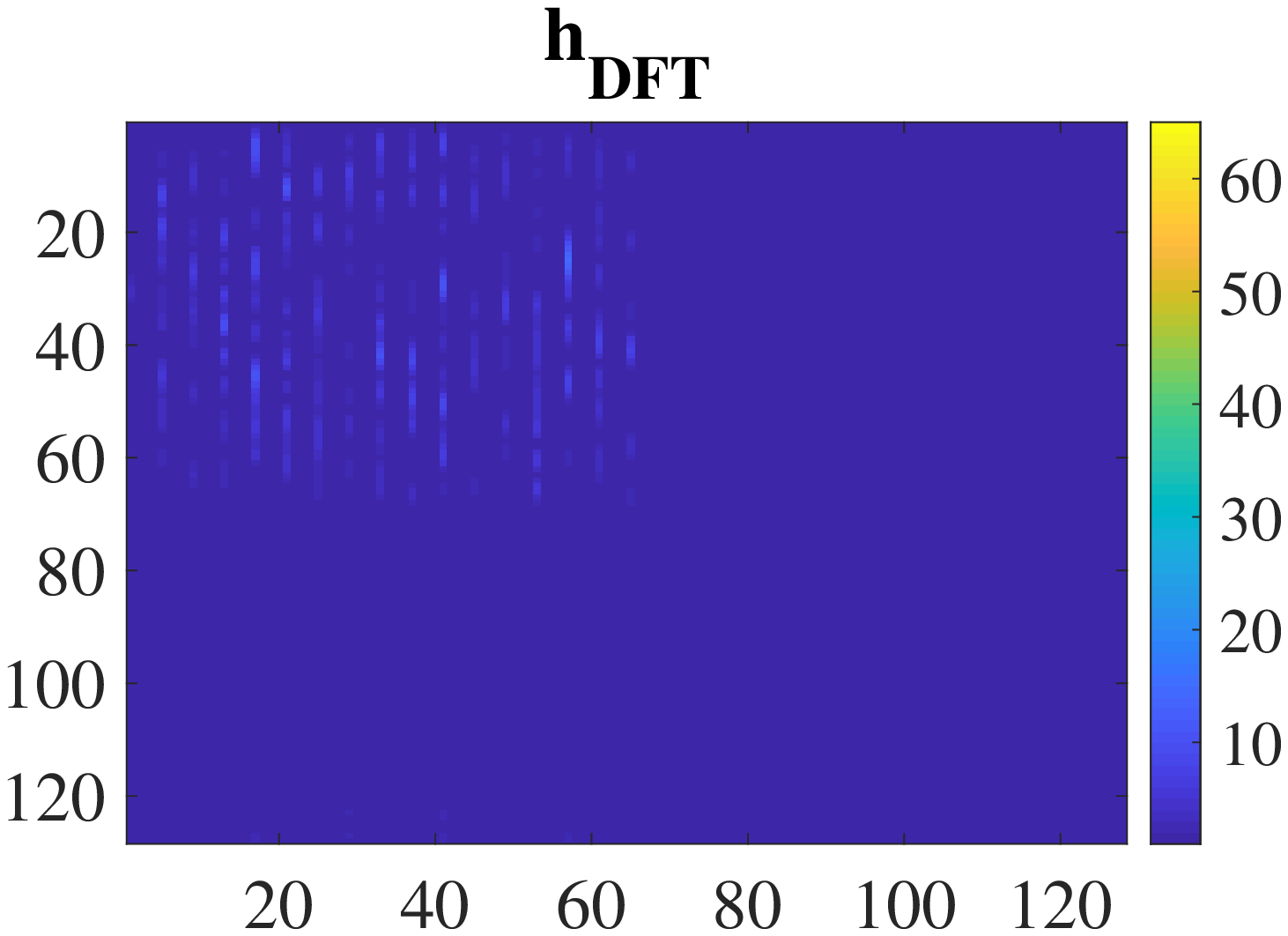} \\
\small (b) 
\end{tabular}
\begin{tabular}{c}
\includegraphics[width=.47\linewidth]{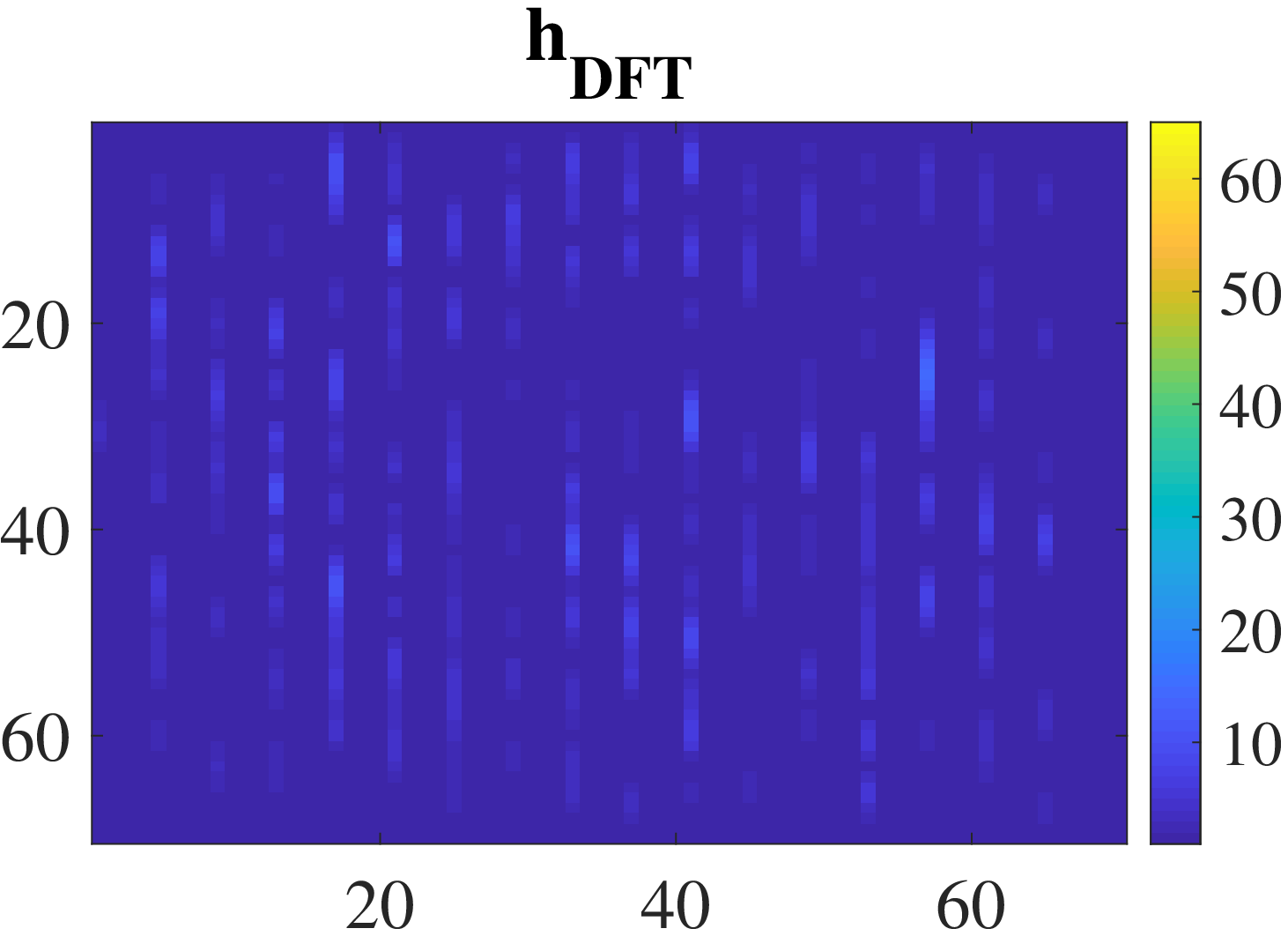} \\
\small (e) 
\end{tabular}
\begin{tabular}{@{}c@{}}
\includegraphics[width=.47\linewidth]{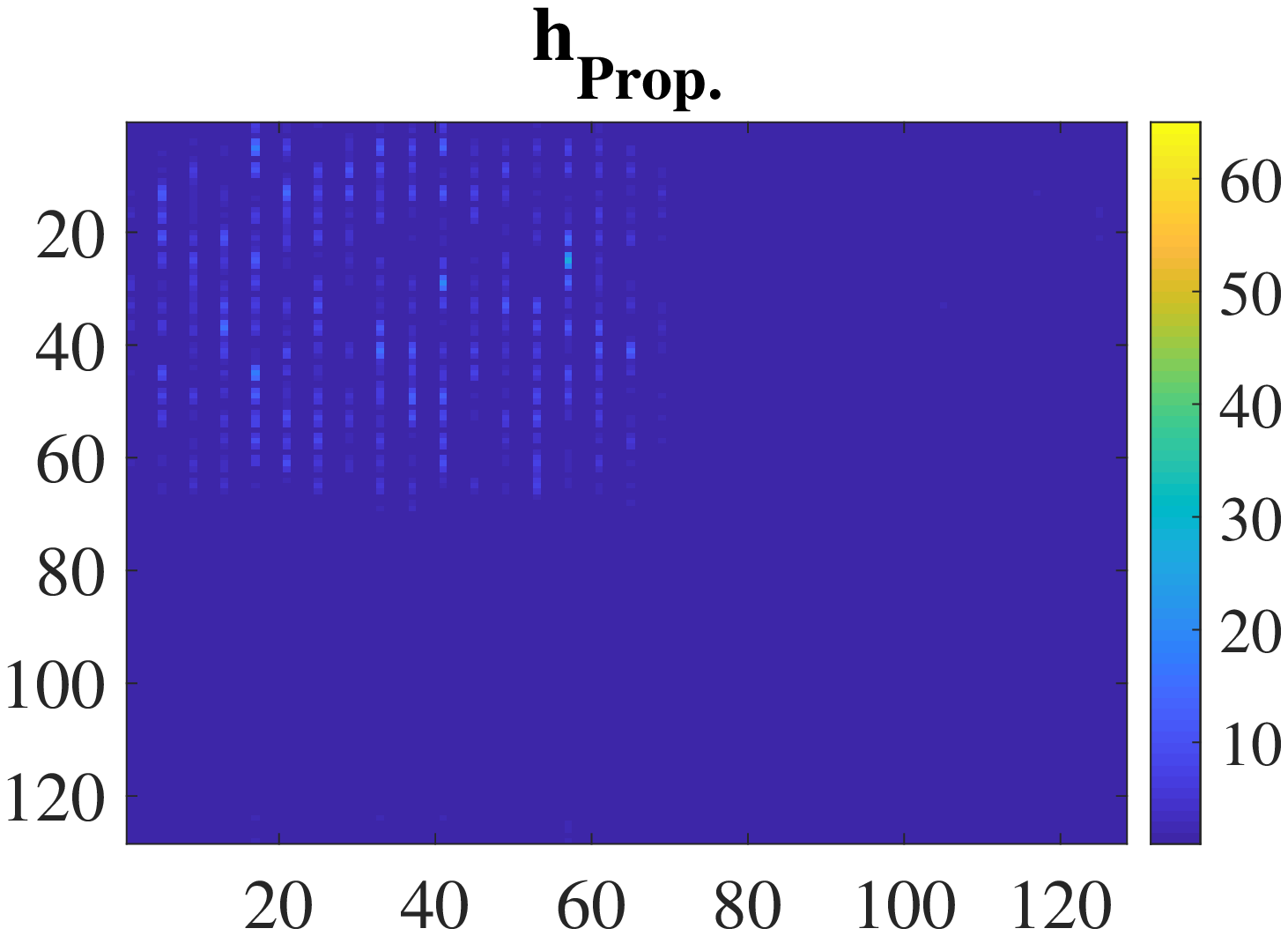} \\
\small (c) 
\end{tabular}
\begin{tabular}{c}
\includegraphics[width=.47\linewidth]{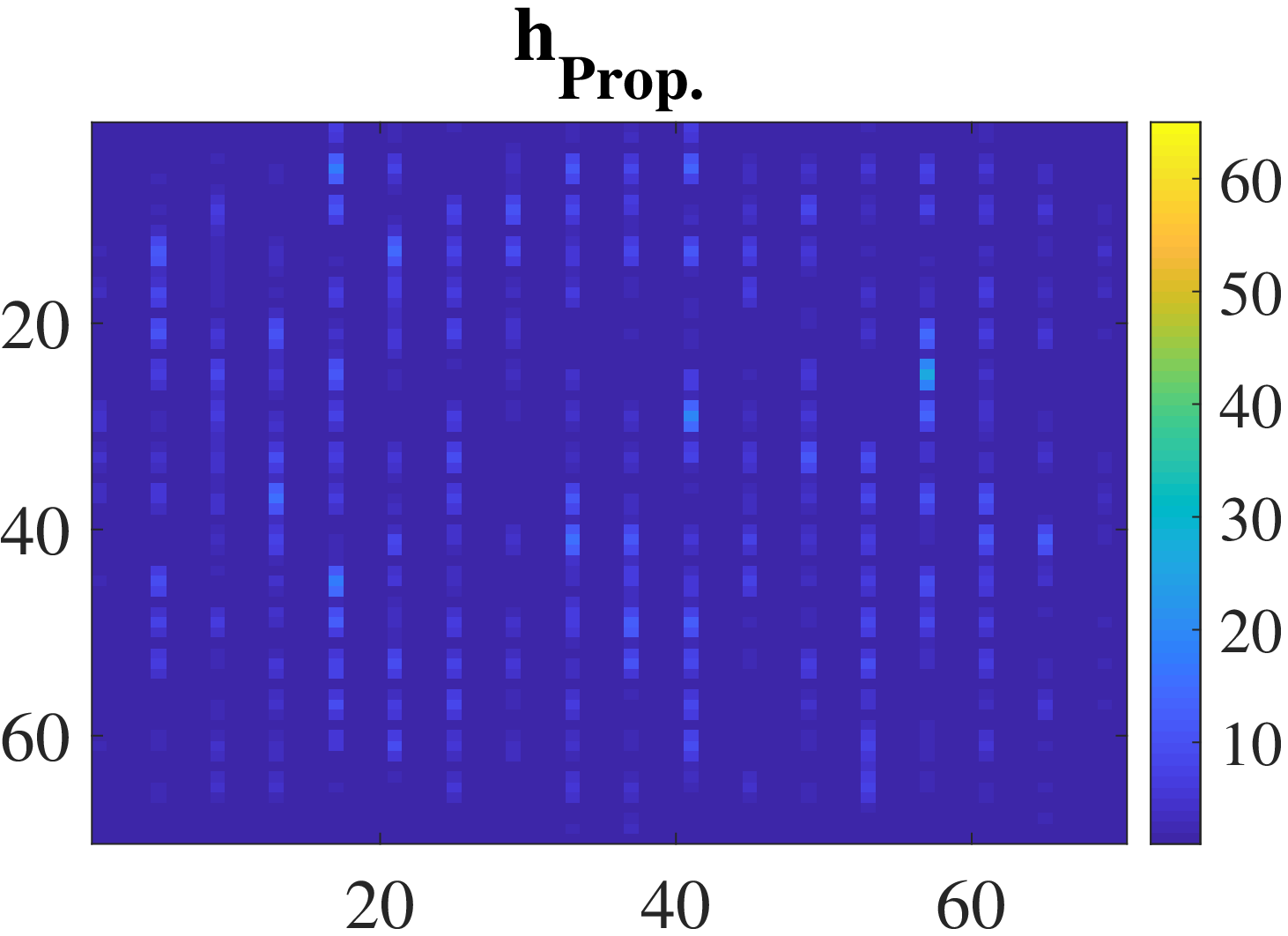} \\
\vspace{-0.2cm}
\small (f) 
\end{tabular}
\caption{Channel sparsity rendering through a DFT and a learned dictionary: true channel, a DFT estimate and a dictionary estimate in (a) through (c), respectively. Corresponding zoomed insets are in (d) through (f), respectively.}
\label{fig555}
\end{figure}

 A learned dictionary can lead to a domain where a signal exhibits multi-dimensional sparsity, where the discrete Fourier transform (DFT) only exposes frequency domain sparsity. The following test empirically supports this idea. In this test, a dictionary is trained over a training set of VCM channel realizations \footnote{A training set can be obtained by either practical measurement campaigns or channel sounding techniques. It is noted that a dictionary is trained for each cell. The dictionary can be updated regularly to account for changes in the far scatterers.} according to the specifications detailed in Section \ref{Secion4}. Then, we examine the time-frequency coefficients of the channel realization. Fig.~\ref{fig555} compares the true CIR to its estimates obtained with sparse coding over a DFT basis and over a learned dictionary where the SNR is -5 $dB$. In view of Fig.~\ref{fig555}, it is seen that a learned dictionary promotes sparsity both in time and frequency. However, this is not the case for the DFT basis, where it produces a sparse representation over the frequency axis, and does not render sparsity over the time axis. 

\par A dictionary initialization of strongly sparse signals in certain domains is a reasonable way to ensure combining sparsity in multiple domains. This stems from the fact that dictionary columns (referred to as atoms) are typically updated by similar training data \cite{mailhe2013fixed}. Thus, dictionary atoms corresponding to sparsity in a certain domain will be updated selectively by training signals of the same sparsity domain. In a channel estimation context, A formalization of the procedure for obtaining a single dictionary addressing multidimensional sparsity is outlined in Algorithm \ref{Algorithm0}. After dictionary internalization accounting for different sparsity aspects, a training set is then used to further tune this dictionary (Steps 2 through 5). Algorithm \ref{Algorithm1} represents the main steps of channel estimation using a dictionary addressing sparsity in angle-delay and Doppler domains and obtained according to Algorithm \ref{Algorithm0}.

\begin{algorithm}[h!]
\caption{Dictionary Learning with Multidimensional Sparsity; in Angle, Delay, and Doppler-Domains.}
\label{Algorithm0}
\begin{algorithmic}[1]{
\renewcommand{\algorithmicrequire}{\textbf{Input:}}
\renewcommand{\algorithmicensure}{\textbf{Output:}}
\REQUIRE A training set of signal realizations $\boldsymbol{Y}$, sets of signal realizations having strong sparsity in angle, delay, and Doppler domains; $\boldsymbol{Y}^a, \boldsymbol{Y}^d$, and $\boldsymbol{Y}^D$, respectively, the intended sparsity level $s$, and the No. of iterations $Num$.
\ENSURE A learned dictionary $\boldsymbol{D}$.
\STATE{Obtain an initial dictionary $\boldsymbol{D}^0$ as
$\boldsymbol{D}^0 \gets [\boldsymbol{Y}^a \boldsymbol{Y}^d \boldsymbol{Y}^D]$, 
initialize $i\gets 0$.}
 \WHILE{$i \leq Num$}
 \STATE{Solve:
$\operatorname*{arg\,min}_{\boldsymbol{W^i}} {\|\boldsymbol{Y}-\boldsymbol{D^iW^i}\|}_2^2 ~ s.t. ~ 
{\|\boldsymbol{W^i}\|}_0 <s$
 }
 \STATE{Update $\boldsymbol{D}^i$ by solving:
$\operatorname*{arg\,min}_{\boldsymbol{D^i}} {\|\boldsymbol{X}-\boldsymbol{D^iW^i}\|}_2^2$
\STATE{Update: $i\gets i+1$.}
 }
 \ENDWHILE
\RETURN $\boldsymbol{D^i}$
}
\end{algorithmic}
\end{algorithm}

\subsection{Dictionary Atom Relation with DFT Columns}

\par
\begin{figure}[!thb]
\centering
\resizebox{0.99\columnwidth}{!}{
\includegraphics{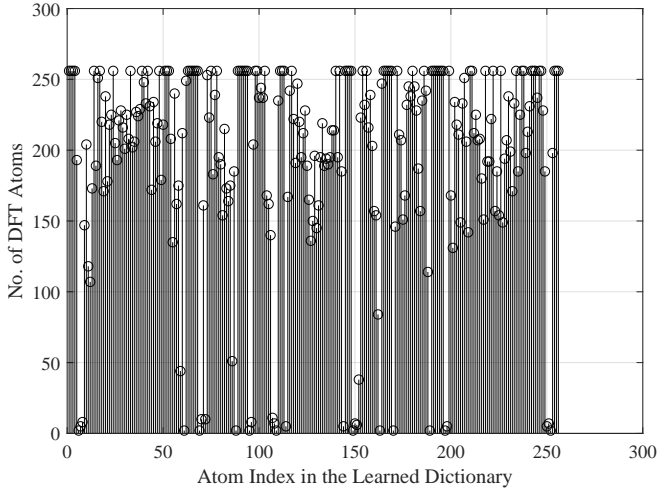}}
\linespread{1}
\caption{Atoms of a learned dictionary correspond to a compositions of columns in a DFT basis matrix.}
\label{motivation}
\end{figure}

 In principle, a dictionary atom is a prototype signal that can be exposed as a linear combination of DFT columns. A well-known advantage of a learned dictionary is its sparser representation \cite{starck2015sparse}\footnote{Other notable advantages are being locally adaptive to the class of training data signals, accounting for measurement impairments, and improving the representation quality \cite{starck2015sparse}.}. Thus, it can be generalized that there is a correspondence between sparsity over a dictionary atom and that over DFT columns. To realize the existence of such a relation, the following test is conducted. A dictionary is trained over 10000 example channel realizations according to the experimental setup provided in Section \ref{Secion4}. A test set of (other) 1400 channel realizations is generated. Then, we quantify the sparsity of each training set over the dictionary and then over a DFT matrix. Next, we plot the correspondence between these sparsity levels in Fig.~\ref{motivation}. The figure shows that a dictionary atom can be represented by multiple DFT columns. Thus, there is a correspondence between sparsity over dictionary atoms and sparsity over DFT columns.

\begin{algorithm}[h!]
\caption{Channel Estimation Exploiting Multidimensional Sparsity}
\label{Algorithm1}
\begin{algorithmic}[1]{
\renewcommand{\algorithmicrequire}{\textbf{Input:}}
\renewcommand{\algorithmicensure}{\textbf{Output:}}
\REQUIRE Training pilots $\boldsymbol{P}$ and their locations $\boldsymbol{I}$, received signal $\boldsymbol{y}$, and a dictionary $\boldsymbol{D}$ rendering sparsity in multiple domains, obtained with Algorithm \ref{Algorithm0}.
\ENSURE A CIR estimate $\hat{\boldsymbol{h}}$.
\STATE{Set pilots symbols at their respective locations in the subcarriers, and transmit the message over the channel to receive $\boldsymbol{y}$.}
\STATE{Apply FFT on $\boldsymbol{y} \xrightarrow{} \boldsymbol{Y}$}
\STATE{Estimate $s$ according to Algorithm \ref{Algorithm2}}
\STATE{Calculate an initial channel estimate:\\ $\boldsymbol{H}_{LS}(\boldsymbol{I})=\boldsymbol{P}^\dagger \boldsymbol{Y}(\boldsymbol{I})$}\footnote{LS stands for the least-squares solution, in the Moore-Penrose pseudo inverse sense.}
\STATE{Solve: 
$\boldsymbol {w}_e=\argminl_ {{\boldsymbol{w}}} {\|\boldsymbol{H}_{LS}-\boldsymbol{D}\boldsymbol{w}\|}_2^2 ~ s.t. ~ {\|\boldsymbol{w}\|}_0 <s $}
\STATE{Obtain a channel estimate:\\
$\hat{\boldsymbol{H}}=\boldsymbol{D}\boldsymbol {w}_e$}
}
\end{algorithmic}
\end{algorithm}

\subsection{Estimating the Unknown Sparsity Level}

\par It is noted that the sparsity level directly depends on the sparsifying transform used. Intuitively, one can roughly estimate the sparsity level of a given signal with respect to a certain sparsifying basis by quantifying the dominant sparse coding coefficients. Since sparsification via fixed basis functions such as the DFT is an inner product process followed by thresholding the representation coefficients, it is convenient to perform sparsity estimation over the DFT. However, sparse coding over a learned dictionary is not achieved through an inner product. Rather, the sparse representation problem reported in (\ref{eq111}). Therefore, it is a sound idea to infer the sparsity level over the dictionary with that over a compact DFT basis. Thus, we propose estimating the sparsity in the dictionary domain in two steps. First, estimating the DFT-domain sparsity. Second, inferring the dictionary domain sparsity level from that of the DFT domain. In this context, the mapping between the two sparsity levels can be inferred using ML techniques.

\par In essence, a dictionary atom is a prototype signal which is richer in content and structure compared to a DFT basis function \cite{ksvd}. To this end, one can write a dictionary atom as a linear combination of many DFT columns. Hence, if a signal is sparse in a dictionary $\boldsymbol{D}$ with sparsity $s$, then it can be expanded as a linear combination of $s$ dictionary atoms as follows
\begin{equation}
 \boldsymbol{y}=\boldsymbol{Dw}=\sum^s_{i=1} \boldsymbol{D}_\lambda(i) a_{i},
\end{equation}
\noindent where $\lambda(i)$ denotes the index of the $i\mbox{-}$th selected atom in the representation of $\boldsymbol{x}$, and $a_{i}$ is a scalar.

\par Each atom in the representation of $\boldsymbol{x}$ can be expanded in terms of many DFT columns as $\boldsymbol{D}_\lambda(i)=\sum_{j=1}^{\sigma_i} \boldsymbol{F}_j$, where $\sigma_i$ denotes the number of nonzero elements in this expansion, and $\boldsymbol{F}$ represents a DFT basis matrix. To this end, it can be seen that the atom $\boldsymbol{D}_\lambda(i)$ corresponds to selecting $\sigma_i$ columns in $\boldsymbol{F}$. {Therefore, the sparsity over $\boldsymbol{D}$ is less than that over $\boldsymbol{F}$, and they are directly related. This proposition is investigated in the analysis detailed in the Appendix.}

\par One can employ an ML model that can be trained to figure out the above-mentioned correspondence between sparsity levels. Once this model is trained, it can be used to estimate the sparsity level. In the training stage, one requires a training set of example signals $\boldsymbol{X}$. For each training vector, the inner product of this vector with a DFT basis is calculated and used as the training feature vector. Also, the sparse coding of this vector over the given dictionary is calculated within a certain error tolerance, and the sparsity of this representation is calculated as the number of nonzero elements and used as the target instances of the model. In the testing stage, for a given observation $\boldsymbol{X}$, one finds its inner product with the DFT basis and feeds this product vector as a feature vector to the ML model. Then, the ML model will predict the corresponding sparsity level of $\boldsymbol{X}$ over $\boldsymbol{D}$.

\par
\begin{figure}[!t]
\centering
\resizebox{0.85\columnwidth}{!}{
\includegraphics{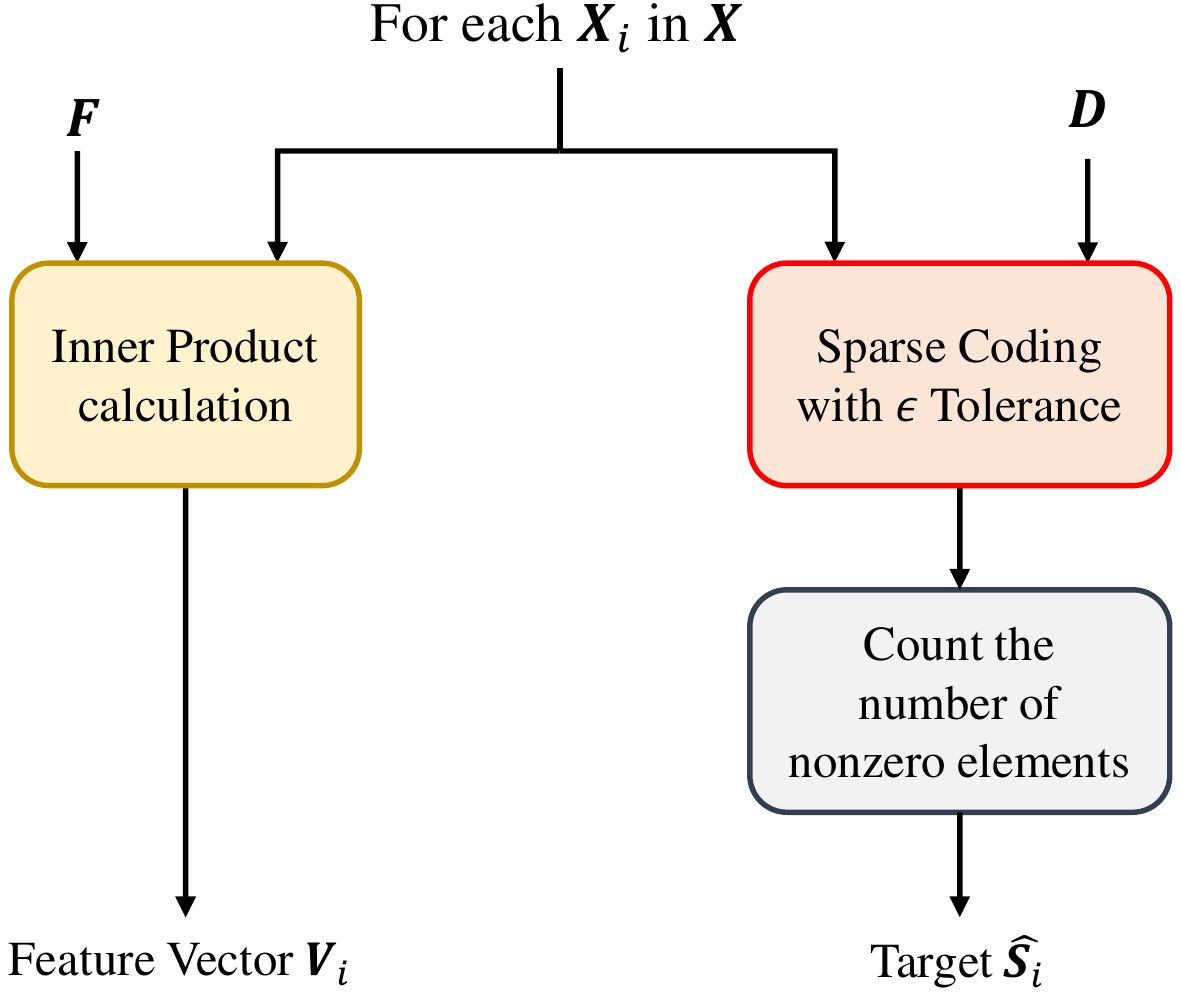}}
\linespread{1}
\caption{The proposed feature extraction for sparsity level estimation.}
\label{fig1000}
\end{figure}

\begin{figure}[!t]
\centering
\begin{tabular}{@{}c@{}}
\includegraphics[width=.33\linewidth]{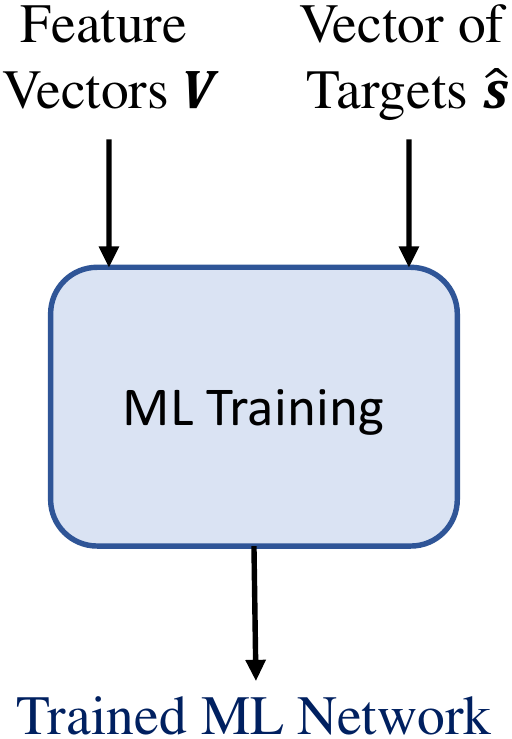} \\
\small (a) 
\end{tabular}
\begin{tabular}{c}
\includegraphics[width=.41\linewidth]{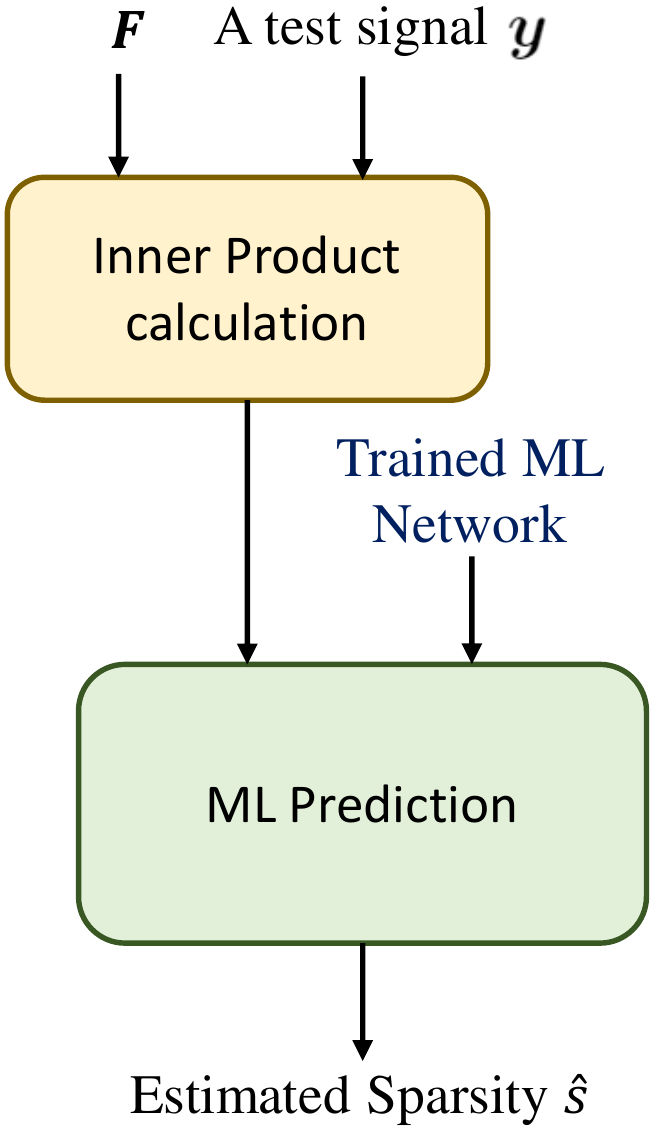} \\
\small (b) 
\end{tabular}
\caption{The proposed ML model for sparsity level estimation; (a) training and (b) testing stages.}
\label{fig2000}
\end{figure}

\begin{figure}[htb]
\centering
\resizebox{0.95\columnwidth}{!}{
\includegraphics[width=14cm]{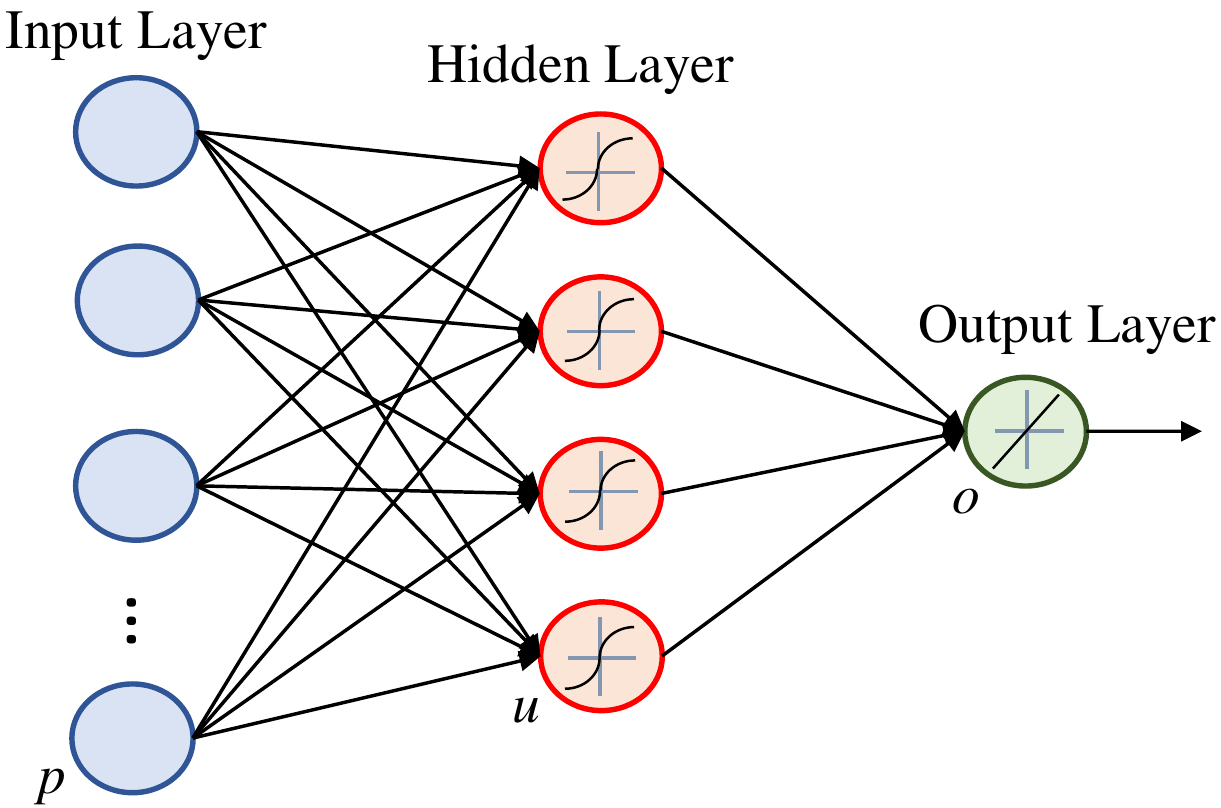}}
\linespread{1}
\caption{The block diagram of the proposed ML model composed of an input layer, a hidden layer, and an output layer.}
\label{nnarch}
\end{figure}

 The above-mentioned feature extraction is pictorially illustrated in Figs. \ref{fig1000} and \ref{fig2000} where $\boldsymbol{V}$ is the matrix of feature vectors, and $\boldsymbol{\hat{S}}$ is the vector of corresponding targets (estimated sparsity levels). Besides, Fig.~\ref{nnarch} demonstrates the overall block diagram for the proposed ML-based algorithm for sparsity level estimation where $p$, $u$, and $o$ represent the number of inputs, hidden layers, and output units, respectively. This model has sigmoid transfer functions in the hidden units and a linear transfer function in the output unit. The main steps of the proposed sparsity level estimation are formally stated in Algorithm \ref{Algorithm2}.

\begin{algorithm}[htb]
\caption{The Proposed Sparsity Level Estimation.}
\label{Algorithm2}
\begin{algorithmic}[1]{
\renewcommand{\algorithmicrequire}{\textbf{Input:}}
\renewcommand{\algorithmicensure}{\textbf{Output:}}
\REQUIRE A signal $\boldsymbol{y}$, a dictionary $\boldsymbol{D}$, a DFT matrix $\boldsymbol{F}$, and a trained ML model $\boldsymbol{f}$.
\ENSURE $\hat{s}$: The estimated sparsity level of $\boldsymbol{y}$ over $\boldsymbol{D}$.
\STATE{Perform an inner product $\boldsymbol{p}=<\boldsymbol{y},\boldsymbol{D}>$}
\STATE{Construct a feature vector $\boldsymbol{v}=[\boldsymbol{p} \boldsymbol{y}]$}
\STATE{Predict $\hat{s}=\boldsymbol{f}(\boldsymbol{D},\boldsymbol{v})$}
\RETURN $\hat{s}$
}
\end{algorithmic}
\end{algorithm}

\subsection{A Note on Computational Complexity}

\par The proposed sparsity level estimation algorithm requires performing an inner product with a DFT basis, thresholding, and then ML prediction. Thus, its computational complexity depends on dictionary learning, sparse recovery, and ML prediction.

\par Let us consider K-SVD \cite{ksvd}, OMP \cite{pati1993orthogonal}, and feedforward ML \cite{ethemalpaydin} as examples of dictionary learning, dictionary learning, sparse recovery, and ML, respectively. The total complexity of K-SVD working on a training set $\boldsymbol{Y}\in \mathbb{C}^{n \times l}$, with sparsity $s$ and $Num$ iterations is $\bigO(Num(s^2+n)kl)$ \cite{skretting2010recursive}.

\par Within the several OMP implementations, let us, for example, consider the naive OMP algorithm working on a signal $\boldsymbol{y} \in\mathbb{C}^n$ over a given dictionary $\boldsymbol{D} \in \mathbb{C}^{n \times k}$, with sparsity $s$. In \cite{sturm2012comparison}, it is shown that the $i\mbox{-}$th iteration's computational complexity is about $\bigO(nk +ks + ks^2 + s^3)$. With sparsity $s$, the overall computational complexity will be $\bigO(nks +ks^2 + ks^3 + s^4)$. Besides, the memory required for the naive approach is $\bigO(nk)$.

\par A feedforward ML architecture has $u(p+1)$ weights in the first layer (hidden layer) and $o(u+1)$ weights in the second layer (output layer). Therefore, its overall testing computational complexity is $\bigO(e.u.(o+p))$ approximately, where $e$ represents the number of training epochs \cite{ethemalpaydin}.

\section{Simulations and Experiments}
\label{Secion4}

\subsection{Parameter Settings}

\par We investigate the performance of the proposed sparsity level estimation algorithm in two case studies. All tests are made with MATLAB R2019a. In both of the cases studies, the MATLAB Neural-Network-Toolbox ``fitting tool" \cite{matlab} is used with a two-layer feed-forward model to map a dataset of numeric inputs and a set of numeric targets. The number of hidden units $u$ is set to 10 while the number of the output units $o$ is one to calculate the sparsity level of the signal. The tan-sigmoid transfer function and linear transfer function are used in the hidden layer and output layer, respectively. Besides, the Random data division function (dividerand), Levenberg-Marquardt training function (trainlm), and mean-square error performance function (mse) are used for data division, training, and as a performance metric, respectively. In this model, the optimum learning rate is found to be 0.001. Finally, the model is trained with 35 and 46 epochs for case study 1 and case study 2, respectively. It is worth mentioning that all hyperparameters are tuned empirically so that they maximize the generalization capability and the performance of the model trained. In this context, generalization means the applicability of the model to successfully work with validation dataset \footnote{The validation dataset is the dataset used in the ML context to provide an unbiased evaluation of a model fit on the training dataset \cite{ethemalpaydin}.}.

\begin{table}[t!]
\centering
\caption{Channel estimation simulation parameters.}
\resizebox{0.9\columnwidth}{!}{
\begin{tabular}{|l|c|}
\hline
\textbf{Property} & \textbf{Value} \\
\hline\hline
Modulation & 16 QAM \\
\hline
\multirow{2}{*}{Channel delay unit} & Sample\\
& Period\\
\hline
Signal length & 100 \\
\hline
Oversampling rate & 10 \\
\hline
Max. no. of resolvable AoAs & 50\\ \hline
Max. no. of resolvable AoDs & 50\\ \hline
Max. no. of resolvable delays & 9\\ \hline
Max. no. of resolvable Doppler &\multirow{3}{*}{9} \\ shifts within the channel spreads & \\ \hline
Sparsity level in & Random no. \\the delay-Doppler domain & between 1 and 10\\ \hline
Sparsity level in & Random no. \\the angular domain & between 1 and 10\\ \hline
Doppler spread & 9\\ \hline
Delay spread & 9\\ \hline
No. of propagation subpaths & \multirow{2}{*}{4}\\ in each resolution bin & \\ \hline
No. of subcarriers & 128\\ \hline
Guard band length & 32\\ \hline
\end{tabular}}
\label{Table2}
\end{table}

\subsection{Performance Evaluation in Channel Sparsity Level Estimation}

\par In this experiment, we consider a MIMO OFDM setting with pilot separation 4. A channel realization is created according to the GSCM and VCM. Specifically, the channel is assumed to be sparse in the angle domain when GSCM is used and sparse in the angle, delay, and Doppler domains when VCM is used. The time axis extends for 16 ms, whereas the frequency axis ranges between 10 GHz and 10.2 GHz. For each received signal realization, different data streams and channel realizations are randomly obtained. For simulating the GSCM, we adopt the experimental setup used in \cite{Ding_Rao}. On the other hand, detailed specifications of the simulation parameters and VCM are listed in Table \ref{Table2} \footnote{It is noted that these are sample arbitrary values set empirically, and the performance does not change with other values.}.

\par As for dictionary learning, a training set of 10$^3$ downlink channel realizations is used to train for the dictionary. We use the K-SVD algorithm \cite{ksvd} for dictionary learning as specified in Algorithm \ref{Algorithm0}, and OMP \cite{pati1993orthogonal} for sparse coding. The dictionary is set to have 256 atoms.

\par
\begin{figure}[bt]
\centering
\resizebox{0.99\columnwidth}{!}{
\includegraphics{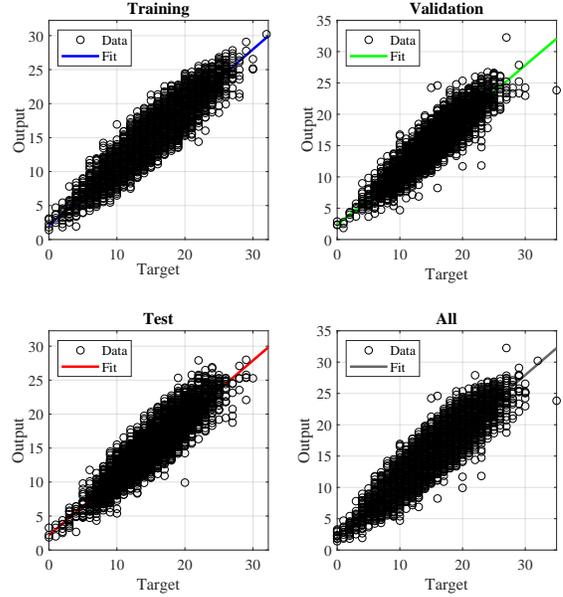}}
\linespread{1}
\vspace{-0.4cm}
\caption{Sparsity level estimation performance in the channel scenario.}
\label{sparsity_level}
\end{figure}

\begin{figure}
\centering
\resizebox{0.99\columnwidth}{!}{
\includegraphics{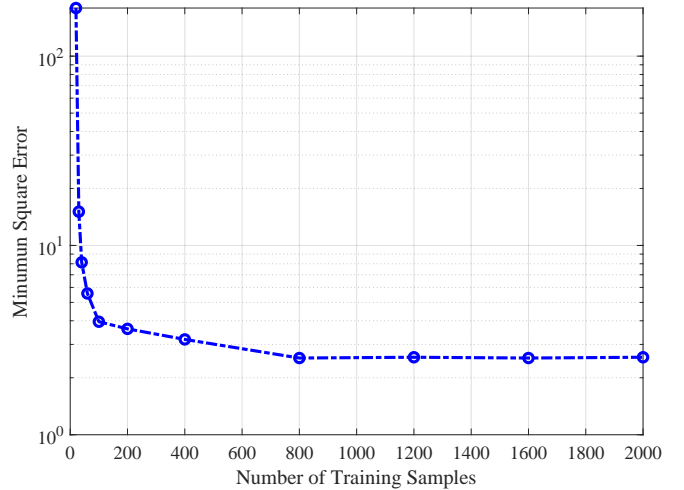}}
\linespread{1}
\caption{Performance of the sparsity level estimation algorithm for different numbers of training samples in the channel scenario.}
\label{variousnumberoftraining}
\end{figure}

 First, we examine the performance of the proposed sparsity level estimation algorithm. For this purpose, we generated GSCM channel realizations where the sparsity level is in the angle domain. Fig.~\ref{sparsity_level} shows the results of this test. In this figure, there are 1400 channel samples (rows), each includes 128 feature columns and one numeric output value for each system and each system includes 16 users. Therefore, 22400 samples are used in total. The dataset is divided into training, validation, and testing with 60\%, 20\%, and 20\% ratios, respectively. It is seen in Fig.~\ref{sparsity_level} that the proposed ML algorithm successfully predicts the sparsity levels as the predicted and the true sparsity levels are very close. This is consistently the case for almost all sparsity levels considered. Also, we do analyses for different numbers of training samples (systems) to learn the optimum number of training samples. Fig.~\ref{variousnumberoftraining} shows that the performance of the proposed algorithm is significantly improved when the number of training pilot symbols increases to 800. However, from 800 to 2000, the improvement is marginal.

\par
\begin{figure}[t]
\centering
\resizebox{0.99\columnwidth}{!}{
\includegraphics{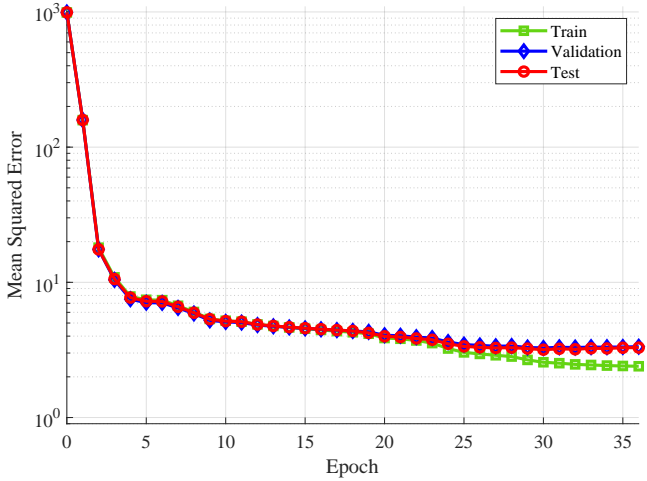}}
\linespread{1}
\vspace{-0.7cm}
\caption{The loss graph of the trained model in the channel scenario.}
\label{loss}
\end{figure}

 As the proposed sparsity estimation algorithm is based on ML, it is important to verify that the established model generalizes well, i.e., does not memorize the inputs during the training stage. To investigate this, the training and validation losses versus epochs for sparsity level estimation are plotted in Fig.~\ref{loss}. In view of this figure, it is evident that the accuracy of the training sets converges to the validation set after $12$ epochs. This means the absence of over-fitting, and shows the generalizability of the proposed model-its ability to work with unforeseen data.

\begin{figure}[t]
\centering
\resizebox{0.99\columnwidth}{!}{
\includegraphics{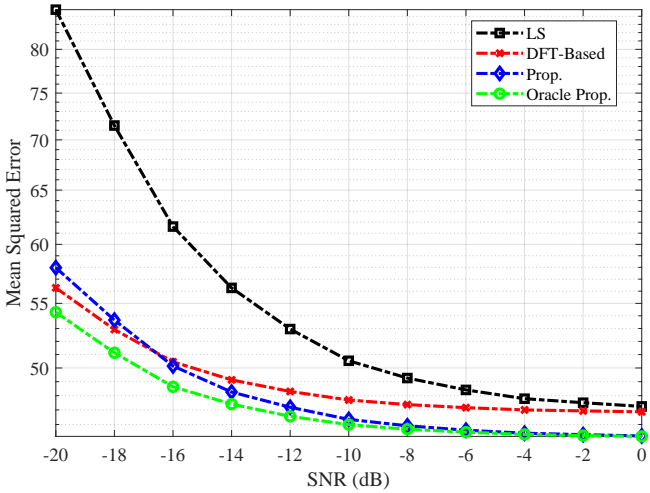}}
\linespread{1}
\vspace{-0.7cm}
\caption{MSE of channel estimation with: LS, sparse coding over a DFT basis with known sparsity level, the proposed algorithm for with estimated sparsity level, and an oracle scenario where the support (nonzero coefficient locations) and sparsity level are known.}
\label{varioussnr}
\end{figure}

\par Now, it is also interesting to test the quality of channel estimation exploiting multi-domain sparsity with its levels obtained by the proposed sparsity level estimation algorithm. For this purpose, we generated VCM channel realizations where the sparsity exists in the angle, delay, and Doppler domains. We compare the performances of the following channel estimation settings.
\begin{itemize}[leftmargin=*]
\item $LS$: LS-based channel estimation, i.e., the solution is based on the knowledge of the pilots and their locations in an inverse problem
\item $DFT$-$based$: sparse recovery over a DFT basis assuming known sparsity levels
\item $Prop.$: multi-domain sparse coding, according to Algorithm \ref{Algorithm1}, with a dictionary learned according to Algorithm \ref{Algorithm0}. Sparsity level is estimated according to the proposed sparsity level estimation algorithm, according to Algorithm \ref{Algorithm2}
\item $Oracle$ $Prop.$: a genie-guided scenario of the proposed algorithm where the exact channel sparsity is assumed known, and the exact locations of the nonzero coefficients are also assumed known. This is included as a comparison benchmark.
\end {itemize}

\par The SNR values are set to range between -20 $dB$ and 0 $dB$ with a step size of 2 $dB$. The average MSEs of the above channel estimation techniques over 1000 channel samples are presented in Fig.~\ref{varioussnr}. In view of Fig.~\ref{varioussnr}, one can make the following conclusions. First, the proposed algorithm outperforms the DFT-based and LS-based channel estimation in high SNR values. However, the DFT-based algorithm is superior to the proposed algorithm in the lower SNR values. This is due to the poor sparsity estimation at very low SNR. Besides, $Oracle$ $Prop.$ is superior to all algorithms. This is consistently the case for all SNR values. Still, the advantage of the proposed algorithm is more notable for moderate and high SNR values.

\subsection{Performance Evaluation for CR Spectrum Sparsity Level Estimation}

\par
\begin{figure}[t!] 
\setlength\abovecaptionskip{-0.1\baselineskip}
\centering\resizebox{0.99\columnwidth}{!}{
\includegraphics{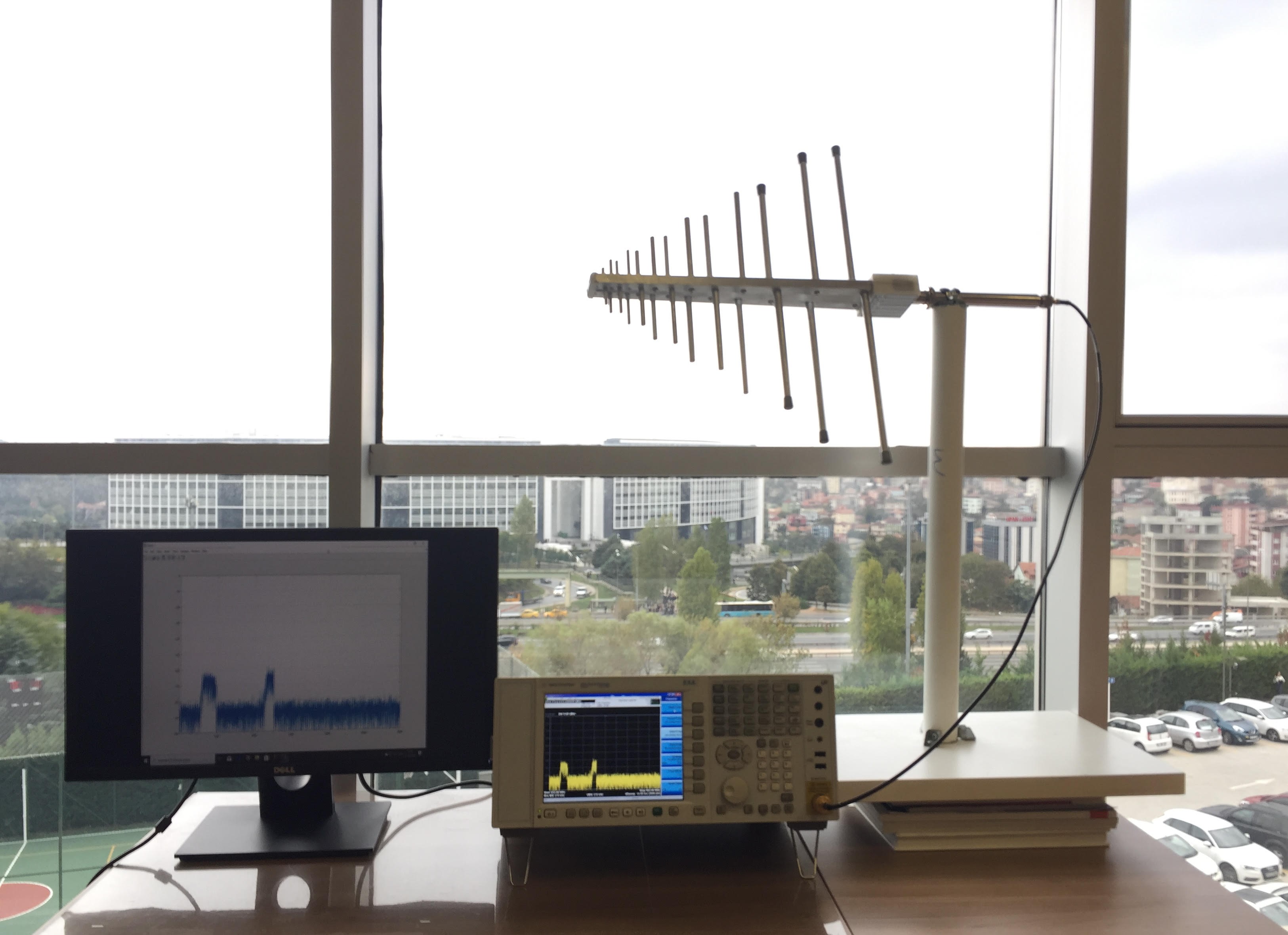}}
\caption{The spectrum measurement setup.}
\label{fig_1}
\end{figure}

 In this test, we consider estimating the sparsity level of the spectrum based on its sampled measurements. Sparsity is rendered over a learned dictionary $\boldsymbol{D}$ that addresses sparsity in both time and frequency domains. Training and testing data are real-world spectrum measurements adopting the experimental setup we used in \cite{aygul2020spectrum}. A picture of this setup is shown in Fig.~\ref{fig_1}. As shown, the UHALP 9108 receiving antenna is connected to the Agilent EXA N9010A spectrum analyzer (SA). The receiving antenna measures received signals between 852-856 MHz with 20 kHz frequency resolution where the frequency-division duplex is considered as an operational mode. The measurement was started at midnight local Istanbul time (GMT+3) on October 16 and measured for five minutes with intervals of one second. This is done using the Keysight Spectrum Analyzer software. Also, data is collected and saved under MATLAB environment. It is worth mentioning that the measurement is conducted in a city surrounded by highways and commercial/residential buildings. Further details on the measurement setup can be found in \cite{aygul2020spectrum}. The dataset includes 60000 samples which are divided as training, validation, and testing with ratios of 60\%, 20\%, and 20\%, respectively.

\par
\begin{figure}[t]
\centering
\resizebox{0.99\columnwidth}{!}{
\includegraphics{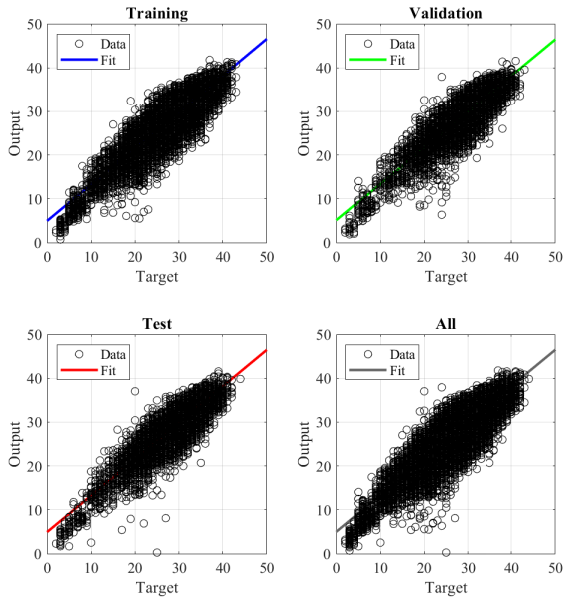}}
\linespread{1}
\vspace{-0.3cm}
\caption{CR spectrum sparsity level estimation performance.}
\label{performance_spec}
\end{figure}

\begin{figure}[t]
\centering
\resizebox{0.99\columnwidth}{!}{
\includegraphics{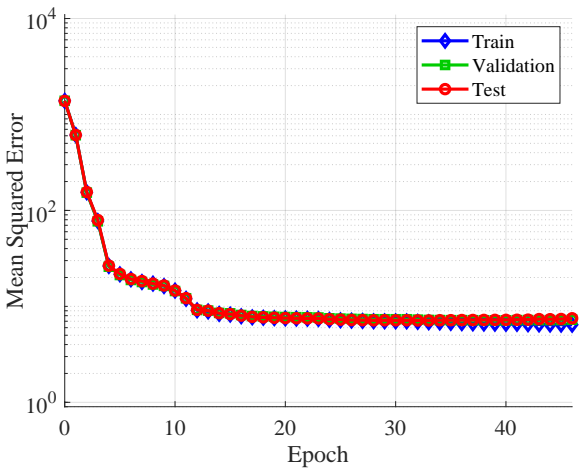}}
\linespread{1}
\vspace{-0.3cm}
\caption{The loss graph of the trained model for CR spectrum sparsity level estimation.}
\label{loss_spectrum2}
\end{figure}

 The experiment is conducted by estimating the sparsity of a set of spectrum measurements, based on the sampled PSD measurements. Fig.~\ref{performance_spec} shows the result of this experiment. It is seen from Fig.~\ref{performance_spec} that the proposed ML algorithm successfully predicts the sparsity levels as the predicted and the true sparsity levels are very close. This is consistently the case for almost all sparsity levels considered. Also, it is verified in Fig.~\ref{loss_spectrum2} that the established model has a good generalization.

\section{Conclusions}
\label{Section5}

\par This paper has shown the possibility of efficiently predicting the sparsity of channels and spectra for applications in 5G and beyond. An ML model has been developed to extract the correspondence between the sparsity level over a given dictionary and that over a square DFT basis. This correspondence allows inferring the level sparsity over the dictionary from its counterpart over the DFT which can be easily calculated. This estimation offers a means by which the CS paradigm can be used in those application areas. This is especially the case when the quantity of interest has sparsity in multiple domains. For this purpose, this paper has presented an algorithm for learning a composite dictionary that renders multi-domain sparsity. By doing so, one can achieve higher degrees of regularization for ill-posed inverse problems in 5G and beyond. Extensive simulations validated the ability of the proposed sparsity level estimation algorithm to successfully and efficiently estimate sparsity in the contexts of channel estimation and spectrum sensing for CR. Besides, it is shown that the proposed dictionary learning algorithm achieves better estimation as it allows for simultaneous sparsity exploitation in multiple domains. Besides, the proposed ML is shown to successfully well-generalize to new unforeseen data.

\appendix

\par The aim is to show that a dictionary atom is a dense prototype signal that can be written as a linear combination of several DFT columns, and that there is a correspondence e between the level of sparsity over a dictionary and that over a DFT basis. First, let us compare the quality of a sparse representation over a DFT basis $\boldsymbol{F} \in \mathbb{C}^{n\times n}$ to that over a redundant (overcomplete) dictionary $\boldsymbol{D} \in \mathbb{C}^{n \times k}$, where $n<k$. Without loss of generality, let us concentrate on time-domain sparsity for simplicity. Herein, the signal of interest is $\boldsymbol{h} \in \mathbb{C}^{n}$. Now, let us compare these representations with a sparsity level of $s$.

\par First, an exact representation of $\boldsymbol{h}$ over $\boldsymbol{F}$ can be obtained using the whole $n$ basis functions (columns) in $\boldsymbol{F}$, as follows
\begin{equation}
 \boldsymbol{h}= \boldsymbol{F}_1 a_1 + \boldsymbol{F}_2 a_2 + \dots + \boldsymbol{F}_n a_n,
\end{equation}
\noindent where $a_1$ through $a_n$ denote the representation coefficients of $\boldsymbol{h}$ with respect to $\boldsymbol{F}$. These can be obtained by performing an inner product between $\boldsymbol{h}$ and $\boldsymbol{F}$.

\par An $s$-sparse representation of $\boldsymbol{h}$ over $\boldsymbol{F}$ can be obtained by selecting the most dominant $s$ coefficients. For simplicity, let us assume that they happen to be the first $s$ coefficients, as follows
\begin{equation}
 \boldsymbol{\hat{h}}_F= \boldsymbol{F}_1 a_1 + \boldsymbol{F}_2 a_2 + \dots + \boldsymbol{F}_s a_s.
\end{equation}

\par Second, with respect to $\boldsymbol{D}$, an $s$-sparse representation of $\boldsymbol{h}$ is:
\begin{equation}
 \boldsymbol{\hat{h}}_D= \boldsymbol{D}_1 b_1 + \boldsymbol{D}_2 b_2 + \dots + \boldsymbol{D}_s b_s.\label{eq100}
\end{equation}

\noindent Again for simplicity, let us assume that the first $s$ atoms (columns) of $\boldsymbol{D}$ are selected, with the corresponding coefficients $b_1$ through $b_s$.

one can assume that it can be expanded spanning many DFT basis functions. So, it can be written as:
\begin{equation}
 \boldsymbol{D}_1= \boldsymbol{F}_1 c_1 + \boldsymbol{F}_2 c_2 + \dots + \boldsymbol{F}_{k} c_{k}.
\end{equation}
\noindent where $k$ is the number of DFT columns required to represent the dictionary atom $\boldsymbol{D}_1$ with coefficients $c_1$ through $c_{k}$. Similarly, the atoms $\boldsymbol{D}_2$ through $\boldsymbol{D}_s$ can be expanded using $k+1$ through $k+s-1$ columns from $F$.

\par Now, (\ref{eq100}) can be rewritten as follows
\begin{equation}
\centering
\begin{split}
\boldsymbol{\hat{h}}_D=& (\boldsymbol{F}_1 c_1 + \dots+ \boldsymbol{F}_{k} c_{k}) b_1 +\\
& \dots +\\
& (\boldsymbol{F}_1 d_1 + \dots+ \boldsymbol{F}_{k} d_{k}) b_s. \\
\end{split}
\end{equation}
\par From the last formulation, it is evident that using the same sparsity level, the sparse representation of $\boldsymbol{h}$ over $\boldsymbol{D}$ is $s$-sparse, in terms of sparsity. However, it is richer in terms of the structure as it is equivalent to using many columns from $\boldsymbol{F}$ \cite{starck2015sparse}. This sets the motivation for assuming that each dictionary atom can be written as a linear combination of DFT columns. Thus, the sparsity of a signal over a dictionary directly corresponds to the sparsity of the same signal over the columns of the DFT basis.

\section*{Acknowledgment}
The work of H. Arslan was supported by the Scientific and Technological Research Council of Turkey (TUBITAK) under Grant 119E433.

\end{document}